# High-Mobility Tri-Gate β-Ga$_2$O$_3$ MESFETs with a Power Figure of Merit over 0.9 GW/cm$^2$

Arkka Bhattacharyya, Saurav Roy, Praneeth Ranga, Carl Peterson, and Sriram Krishnamoorthy

*Abstract*— In this letter, fin-shape tri-gate β-Ga$_2$O$_3$ lateral MESFETs are demonstrated with a high power figure of merit (PFOM) of 0.95 GW/cm$^2$ – a record high for any β-Ga$_2$O$_3$ transistor to date. A low-temperature un-doped buffer-channel stack design is developed which demonstrates record high Hall and drift electron mobilities in doped β-Ga$_2$O$_3$ channels allowing for low ON resistances (R$_{ON}$) in β-Ga$_2$O$_3$ MESFETs. Fin-widths (W$_{fin}$) were 1.2-1.5 μm and there were 25 fins (N$_{fin}$) per device with a trench depth of ~ 1μm. A β-Ga$_2$O$_3$ MESFET with a source-drain length of 6.4 μm exhibits a high ON current (187 mA/mm), low R$_{ON}$ (20.5 Ω.mm) and a high average breakdown field (4.2 MV/cm). All devices show very low reverse leakage until catastrophic breakdown for breakdown voltages (V$_{BR}$) scaled from 1.1kV to ~3kV. This work demonstrates the potential of channel engineering in improving β-Ga$_2$O$_3$ device performance toward lower conduction losses for low-to-medium voltage applications.

## I. Introduction

Ultra-wide bandgap (UWBG) β-Ga$_2$O$_3$ (E$_g$ = 4.6 - 4.9 eV) material and device technology is maturing rapidly and offers enormous opportunities for next-generation solid-state power switching with improved system-level size, weight, and power (SWaP) efficiency. β-Ga$_2$O$_3$ is the only UWBG semiconductor that offers the advantage of producing large area native bulk substrates from melt-grown techniques – offering potentially lowered costs for large-scale manufacturing at a much higher device yield and uniformity[1],[2]. Thanks to its compatibility with the established WBG process technology and single crystal growth using standard epitaxial techniques, β-Ga$_2$O$_3$ material and device performance has improved rapidly with lateral and vertical β-Ga$_2$O$_3$ devices demonstrating class-leading blocking voltages (up to 8kV) and breakdown field strengths (>5 MV/cm)[3]–[10].

Although β-Ga$_2$O$_3$ devices have demonstrated tremendous performance advantages, its performance is still far from its projected intrinsic material limit. β-Ga$_2$O$_3$ transistors with high breakdown voltages and PFOMs have been realized which have focused mainly on electric field management techniques for improving average breakdown fields and device scaling for improving ON resistances[3], [5], [11]–[13]. However, less attention has been paid to doped channel design and engineering for improving electron mobility toward lowered device ON resistance[14]–[16]. In this letter, we demonstrate an improved channel stack design with low-temperature metalorganic vapor phase epitaxy (MOVPE) grown un-doped buffer layers with record high Hall and drift electron mobilities in doped β-Ga$_2$O$_3$ channels. By fabricating fin-shape tri-gate β-Ga$_2$O$_3$ MESFETs, PFOM close to 1 GW/cm$^2$ and multi-kV breakdown voltages (over 2kV) are achieved simultaneously.

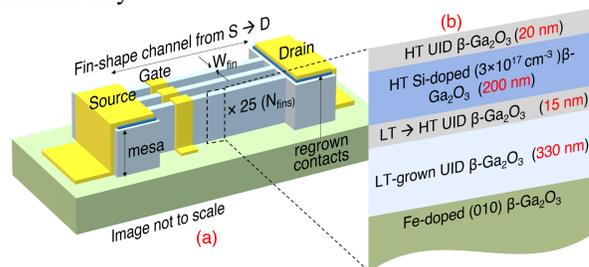

Fig. 1. (a) 3D schematic of the tri-gate β-Ga$_2$O$_3$ MESFETs (with a SiN$_X$ wrap-around passivation not shown). (b) 2D cross-section schematic of the channel stack.

## II. Device Growth and Fabrication

For the channel design, a hybrid low temperature - high temperature (LT-HT) undoped buffer/doped channel epitaxial stack is grown using MOVPE. The (010) Fe-doped Ga$_2$O$_3$ substrates (NCT, Japan) were cleaned in HF (49%) for 30 mins prior to channel growth [17]. The epitaxial structure was grown using an Agnitron Agilis MOVPE reactor with TEGa, O$_2$ and silane (SiH$_4$) as precursors and argon as carrier gas. An LT (600°C) undoped Ga$_2$O$_3$ buffer (330 nm thick) is grown followed by transition layers to a HT (810°C) Si-doped Ga$_2$O$_3$ channel layers (~200 nm) without growth interruption[18].

Fin-shape MESFETs were fabricated with a channel stack whose Hall mobility, sheet charge (electron density) and R$_{sh}$ were 168 cm$^2$/Vs, 5.8×10$^{12}$ cm$^{-2}$ (~3×10$^{17}$ cm$^{-3}$), and 6.4 kΩ/□, respectively. The channel stack cross-section schematic is shown in Figure 1(b). Fin-shape channels with length (L$_{fin}$) that run from the source to the drain (L$_{fin}$ ~ L$_{SD}$) were formed using Ni/SiO$_2$ mask pattern and SF$_6$-Ar ICP-RIE dry etch with trench depths ~ 1 μm (3D schematic shown in Figure 1(a)) [19]. After the dry etching step, wet acid treatments using room temperature diluted HCl (20 mins) and diluted HF (10 mins) were used for dry-etch-induced surface damage recovery. Planar LT-MOVPE regrown ohmic contacts were employed [12]. The estimated electron concentration in the regrown contact layer is around n ~ 1.4 × 10$^{20}$ cm$^{-3}$. Ti/Au/Ni (20/150/50 nm) ohmic metal was evaporated on the regrown contacts and annealed at 450°C for 1.5 mins in N$_2$ ambient[12]. Ni/Au/Ni (30/100/30 nm) gates were evaporated

This material is based upon work supported by the II-VI foundation Block Gift Program 2020-2022. This material is also based upon work supported by the Air Force Office of Scientific Research under award number FA9550-21-0078 (Program Manager: Dr. Ali Sayir).

A. Bhattacharyya & P. Ranga are with the Department of Electrical and Computer Engineering, University of Utah, Salt Lake City, Utah, USA 84112.
S. Roy, C. Peterson and S. Krishnamoorthy are with Materials Department, University of California, Santa Barbara, California, USA 93106. (email: sriramkrishnamoorthy@ucsb.edu).
*corresponding author email: a.bhattacharyya@utah.edu.

to form the Schottky gates. The whole device was passivated using a ~250 nm PECVD (300°C) deposited SiN$_x$ dielectric. Lateral device dimensions were verified by SEM. Fin-widths (W$_{fin}$) were 1.2-1.5 μm, trench widths were ~5.3 μm and there were 25 fins (N$_{fin}$) per device. The L$_{GS}$ and L$_G$ were fixed at 2.4 μm and 1.3 μm and the L$_{GD}$ was varied from 2.7 to 16.7 μm on the same wafer. Concentric Schottky gate CV pads (220 μm diameter) and fatFETs structures were also fabricated on the same MESFET sample.

### III. RESULTS AND DISCUSSIONS

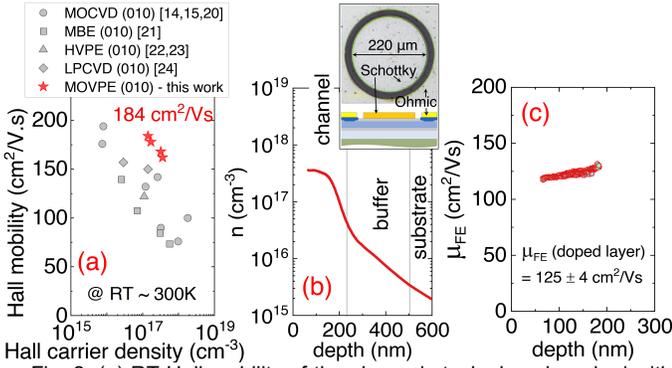

Fig. 2. (a) RT Hall mobility of the channel stacks benchmarked with various state-of-the-art reports [14], [15], [20]–[24]. (b) & (c) channel charge profile extracted from C-V measurements (inset: CV pad image and cross-section schematic) and RT field-effect mobility depth profile in the doped β-Ga$_2$O$_3$ MESFET channel, respectively.

From room temperature (RT) Hall measurements, this stack design is shown to have an effective RT Hall mobility value in the range 162 – 184 cm$^2$/Vs for doped channel electron densities of 1.5 - 3.5 ×10$^{17}$ cm$^{-3}$ measured on multiple samples/substrates. It is hypothesized that the enhanced channel mobility could be due to lower Fe riding into the channel from the substrate due to the lower growth temperature of the buffer as well as the absence of any low-mobility parasitic channel near the substrate [25], [26]. Further details and characterization of this stack design will be reported elsewhere. These Hall mobility values are record high for any doped Ga$_2$O$_3$ channel to date as it is compared with the state-of-the-art values reported utilizing various growth techniques in Figure 2(a).

Charge profile of the MESFET channel extracted from capacitance-voltage measurement is shown in Figure 2(b). It shows that the buffer is completely depleted and there is no active parasitic channel at the epilayer-substrate interface. These further supports ascribing the measured Hall and drift mobility to only the doped channel layer. The channel mobility of the MESFET was characterized using fatFET devices (L$_G$ ~103 μm, L$_{GS}$/L$_{GD}$ ~1 μm) in the linear region of the device operation. Under a low drain bias (V$_{DS}$ = 0.1 V), the field-effect mobility μ$_{FE}$ can be related to the transconductance as, μ$_{FE}$ =(g$_m$×L$_G^2$)/(C$_G$ ×V$_{DS}$) where g$_m$, L$_G$, C$_G$, V$_{DS}$ are the transconductance, gate length, gate-to-channel capacitance and applied drain bias respectively. Figure 2(c) shows the room-temperature depth profile of the extracted μ$_{FE}$ in the doped channel. The μ$_{FE}$ showed an average value of ~125 cm$^2$/Vs with a peak value of 132 cm$^2$/Vs which is the highest electron drift mobility value ever reported in a uniformly doped β-Ga$_2$O$_3$ channels.

Figure 3(a) and 3(b) shows the DC output and transfer curves for the fin-shape MESFET with dimensions L$_{GS}$/L$_G$/L$_{GD}$ = 2.4/1.3/2.7 μm. The ON current and resistance were normalized to the device width (W$_{fin}$ × N$_{fin}$). The devices show clear current saturation and low saturation voltages (V$_{DS,Sat}$). The maximum ON current measured was 187 mA/mm. The ON resistance (R$_{ON}$) extracted from the linear region of the output curves were found to be 20.5 Ω.mm. From TLM measurements, the total R$_C$ (contact resistance) to the channel was extracted to be 1.0 ± 0.2 Ω.mm (≤ 5% of the total device R$_{ON}$). The devices show sharp pinch off with low sub threshold swing (156 mV/dec) and threshold voltage of -10 V. From the transfer curves, the devices show low leakage (~10$^{-11}$A/mm) and high I$_{ON}$/I$_{OFF}$ ratio ~10$^{10}$. The transistors also exhibit very low gate leakage and high transconductance peak of 12.6 mS/mm. The hysteresis effects seem to be minimal as shown in Figure 3(c) and 3(d) dual sweep I-V curves. These devices exhibit a negligible hysteresis of ΔV ~ 0.06 V. Dynamic performance characterization is required in the future to ascertain any deleterious effect of low temperature buffers and any resultant charge trapping.

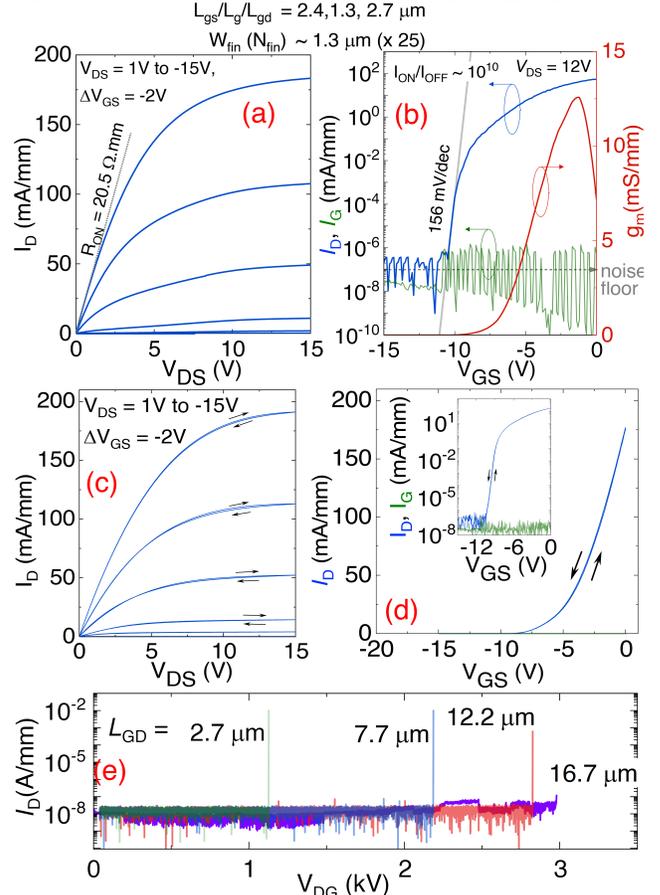

Fig.3. DC (a) output and (b) transfer curves of the tri-gate β-Ga$_2$O$_3$ MESFETs. Dual sweep linear DC (c) output and (d) transfer I-V hysteresis curves (inset: semilog plot) of a device with the same dimensions. (e) Off-state breakdown characteristics of the tri-gate β-Ga$_2$O$_3$ MESFETs with various L$_{GD}$ values.

Figure 3(e) shows the three-terminal breakdown characteristics (at V$_{GS}$ = -35V) of the MESFETs with various L$_{GD}$ values. All the breakdown measurements were performed with the wafer submerged in FC-40 Fluorinert dielectric liquid. The device breakdown was catastrophic (destructive) with negligible reverse leakage until the breakdown occurred. The reverse leakage during the breakdown measurements were limited by the measurement tool (Keysight B1505 with 3kV HV SMU). It is seen in Figure 3(c) that the $V_{BR}$ (= $V_{DS}$-$V_{GS}$) scaled from 1.1kV to ~3kV as the $L_{GD}$ was scaled from 2.7 to 16.7 μm.

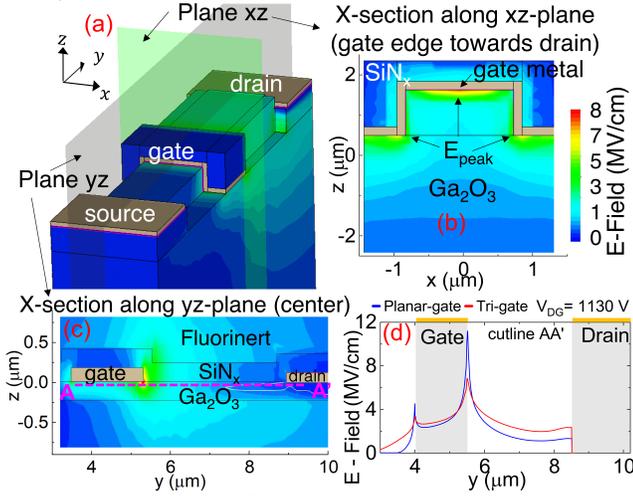

Fig.4 (a) Simulated 3D (SiN$_x$ & Flourinert partly hidden for visibility) (b)&(c) X-section E-field contours along two planes of the tri-gate design with $L_{GD}$ = 3μm, and $V_{DG}$ = 1130V. (d) E-field profile of the same FET (cutline AA' under the gate) comparing planar-gate and tri-gate.

From the cross-section (xz-plane Figure 1(a,b)) of the 3D TCAD simulated structure (with $L_{GD}$ = 3μm, and $V_{DG}$ = 1130V) at the gate edge towards drain, it is shown that the peak fields are at the center of the top gate and gate metal corners in the etched region. Hence, in the presence of the UID cap layer and deeper trenches, peak fields are present in the UID cap region and the insulating substrate. When this design is compared with a planar gate structure (Figure 4(d)), the tri-gate exhibits lower peak field at the gate edge (without the need for field plates), improving the $E_{AVG}$ values in the drift region dramatically.

Figure 5(a) shows the variation of $V_{BR}$ and the effective average field ($E_{AVG}$ = $V_{BR}$/$L_{GD}$) as a function of $L_{GD}$. The maximum $E_{AVG}$ achieved was ~ 4.2 MV/cm for the smallest device with $L_{GD}$ of 2.7 um ($V_{BR}$ = 1.13 kV). This is the highest reported average breakdown field for $L_{GD}$ > 1 μm [3], [27]. The $E_{AVG}$ decreased monotonously as the $V_{BR}$ increased with increasing $L_{GD}$. For a $V_{BR}$ of ~3 kV and $L_{GD}$ of 16.7 μm, the $E_{AVG}$ is 1.8 MV/cm. Since this is the first report of LT MOVPE-grown buffers in β-Ga$_2$O$_3$ devices, further study combining field plates and buffer breakdown structures will be necessary to elucidate on factors limiting the breakdown performance in these devices, especially at higher $L_{GD}$. Nevertheless, the high ON currents (low $R_{ON}$), the high $V_{BR}$, $E_{AVG}$ and low reverse leakage behavior simultaneously demonstrated in these first-generation LT-buffer devices are a significant improvement over the state-of-the-art β-Ga$_2$O$_3$-based transistors.

Since these devices utilized fin lengths running from source to drain, an effective channel area of ($L_{SD}$ + 2$L_T$)×$W_{fin}$×$N_{fin}$ was used to normalize the ON resistance. $L_T$, the transfer length, is extracted from TLM structures to be 0.2 μm. The PFOM values for fin-shape MESFETs are plotted as a function of $L_{GD}$ (Fig.5(b)). The highest PFOM of 0.95 GW/cm$^2$ was calculated for MESFETs with $L_{GD}$ of 2.8 μm and 7.7 μm which had $V_{BR}$ of 1.1 kV and 2.2 kV, respectively. The PFOM of the devices with $L_{GD}$ of 12.2 μm and 16.7 μm were 0.65 GW/cm$^2$ ($V_{BR}$ = 2.8kV) and 0.44 GW/cm$^2$ ($V_{BR}$ ~3 kV). The PFOM of the large $L_{GD}$ devices were a bit lower due to lower $E_{AVG}$ values compared to the small $L_{GD}$ devices, as discussed earlier. Nevertheless, the PFOM reported for > 2 kV devices are still the highest reported values. Figure 5(c) benchmarks the PFOM values of the fin-shape MESFETs with the existing literature reports. It is compared with state-of-the-art β-Ga$_2$O$_3$–based transistors that include advanced designs like Ga$_2$O$_3$ MOSFETs, (AlGa)$_2$O$_3$/Ga$_2$O$_3$ HFETs and p-n hetero-junction β-Ga$_2$O$_3$ FETs. It shows the reported PFOM of 0.95 GW/cm$^2$ is a record high for any β-Ga$_2$O$_3$ transistor to date. Further improvement can be expected by implementing E-field management techniques like field-plates, planar access regions, gate dielectrics with high breakdown field strengths, higher channel charge, higher channel mobility and lowering reverse leakage simultaneously.

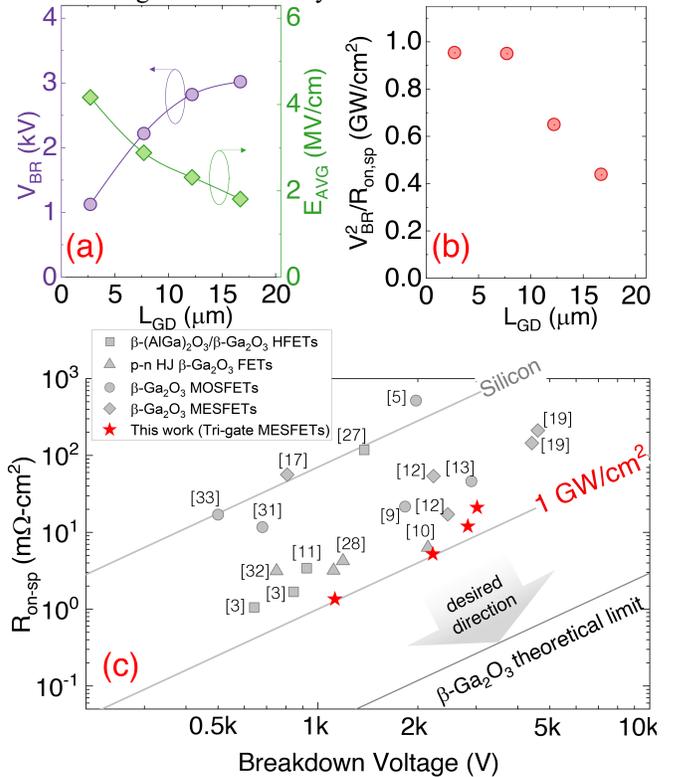

Fig.5 (a) $V_{BR}$ and $E_{AVG}$, (b) PFOM as a function of $L_{GD}$. (c) Differential $R_{on,sp}$-$V_{BR}$ benchmark plot of our Tri-gate β-Ga$_2$O$_3$ MESFETs with the literature reports [3], [4], [4], [5], [9]–[11], [13], [28]–[33].

## IV. CONCLUSION

We demonstrate over 0.9 GW/cm$^2$ PFOM in multi-kV fin-shape β-Ga$_2$O$_3$ lateral MESFETs – a record high for any β-Ga$_2$O$_3$ transistor to date. An LT-HT buffer-channel stack is demonstrated using MOVPE with record high RT Hall and drift mobilities in doped β-Ga$_2$O$_3$ channels. Using tri-gates, β-Ga$_2$O$_3$ MESFETs with high ON currents, negligible hysteresis effects and low ON resistances are realized with very low reverse leakage and $V_{BR}$ values of 1.1kV to ~3 kV. These devices show great potential of high-performance β-Ga$_2$O$_3$ FETs for future power device applications in the low to medium voltage range.